\documentclass[prd,twocolumn]{revtex4-1}

\usepackage{amsmath}
\usepackage{blkarray}
\usepackage{multirow}
\usepackage{mathtools}
\usepackage{graphicx}

\usepackage{graphics}
\usepackage{graphicx}
\usepackage{epsfig}
\usepackage{color}
\begin{document}

\newcommand{\BayesWave}{\texttt{BayesWave}}
\newcommand{\BayesLine}{\texttt{BayesLine}}
\newcommand{\LALInference}{\texttt{LALInference}}
\newcommand{\cWB}{\texttt{cWB }}
\newcommand{\Msun}{{\mathrm{M}_\odot}}
\newcommand{\Mc}{{\mathcal{M}}}
\newcommand{\data}{d}
\newcommand{\h}{{\bm{h}}}
\newcommand{\n}{{\bm{n}}}
\newcommand{\params}{{\boldsymbol \theta}}
\newcommand{\intp}{{\boldsymbol \lambda}}
\newcommand{\extp}{{\boldsymbol \Omega}}
\newcommand{\Hyp}{{\mathcal{H}}}
\newcommand{\Bay}{{\mathcal{B}}}
\newcommand{\Sig}{{\mathcal{S}}}
\newcommand{\Gli}{{\mathcal{G}}}
\newcommand{\Noi}{{\mathcal{N}}}
\newcommand{\Odd}{{\mathcal{O}}}
\newcommand{\IFO}{{\rm IFO}}

\newcommand\Neil[1]{\textcolor{red}{[#1]}}
\newcommand\Tyson[1]{\textcolor{blue}{[#1]}}
\newcommand\Jonah[1]{\textcolor{green}{[#1]}}

\DeclareGraphicsExtensions{.pdf,.gif,.jpg}

\title{Enabling high confidence detections of gravitational-wave bursts}
\author{Tyson B. Littenberg}
\affiliation{Center for Interdisciplinary Exploration and Research in Astrophysics (CIERA) \& Department of Physics and Astronomy, Northwestern University, 2145 Sheridan Road, Evanston, IL 60208}
\email{tyson.littenberg@ligo.org}
\author{Jonah B. Kanner}
\affiliation{LIGO Laboratory, California Institute of Technology, Pasadena, CA 91125, USA}
\author{Neil J. Cornish}
\affiliation{Department of Physics, Montana State University, Bozeman, MT 59717, USA}
\author{Margaret Millhouse}
\affiliation{Department of Physics, Montana State University, Bozeman, MT 59717, USA}

\begin{abstract}
With the advanced LIGO and Virgo detectors taking observations the detection of gravitational waves is expected within the next few years.
Extracting astrophysical information from gravitational wave detections is a well-posed problem and thoroughly studied when detailed models for the waveforms are available. 
However, one motivation for the field of gravitational wave astronomy is the potential for new discoveries.  
Recognizing and characterizing unanticipated signals requires data analysis techniques which do not depend on theoretical predictions for the gravitational waveform.
Past searches for short-duration un-modeled gravitational wave signals have been hampered by transient noise artifacts, or ``glitches,'' in the detectors.  
 In some cases, even high signal-to-noise simulated astrophysical signals have proven difficult to distinguish from glitches, so that essentially any plausible signal could be detected with at most 2-3 $\sigma$ level confidence.  
We have put forth the \BayesWave{} algorithm to differentiate between generic gravitational wave transients and glitches, and to provide robust waveform reconstruction and characterization of the astrophysical signals.  
Here we study \texttt{BayesWave}'s capabilities for rejecting glitches while assigning high confidence to detection candidates  through analytic approximations to the Bayesian evidence.  Analytic results are tested with numerical experiments by adding simulated gravitational wave transient signals to LIGO data collected between 2009 and 2010 and found to be in good agreement.
\end{abstract}

\maketitle

\section{Introduction}

When the LIGO~\cite{LIGO} and Virgo~\cite{Virgo} observatories make their first detection of gravitational waves (GW) it will represent a major achievement. 
Making a claim of a significant discovery requires exceptional evidence.  
In the field of particle physics, a common practice for declaring detection of a new particle is a ``5-sigma'' level of confidence, meaning that there is probability of less than $3 \times 10^{-7}$ of the observation arising from sources other than the claimed discovery.

Detailed theoretical predictions for the gravitational wave signal helps reduce the false alarm (or false positive) rate due to glitches \citep{Abbott:2007kv,Blackburn:2008ah,Abadie:2010yb,Aasi:2012wd} but searches for generic transient signals, known as GW bursts, have been hampered by an inability to distinguish non-Gaussian noise artifacts, or ``glitches,'' and astrophysical signals at high confidence (e.g. Ref.~\citep{Abbott:2009up}).  
Background distributions for burst searches, determined by time-shifting the data from multiple detectors so that no coherent gravitational signals are in the data, show a long tail to high signal to noise ratio (SNR), meaning that even a very strong gravitational wave signal would be consistent with having arisen from a glitch.
In both the first and second joint LIGO-Virgo observation runs, simulated signals intentionally added to the data and included in the final analysis were recovered by Burst algorithms with false alarm probabilities of order 10\% and could not be identified as statistically significant events without using physically motivated waveform models~\citep{Abbott:2009zi, Abadie:2012rq}. 

In preparation for the advanced detector era several new approaches to the burst detection problem have been developed.
Thrane and Coughlin~\cite{Thrane:2015psa} have demonstrated the capability to make high-confidence detections of long-duration  ($\mathcal{O}(10) s$) burst signals in non-stationary, non-Gaussian noise by searching for excess power found along parameterized curves through a time-frequency representation of the data.
In an independent effort, the Bayesian parameter estimation analysis library \LALInference{}~\cite{LALInference}, originally designed for the characterization of compact binary signals, has been adapted for burst analyses by using a sine-Gaussian waveform as the gravitational wave template~\cite{Essick:2014wwa,oLIB}.  \LALInference{} differentiates between signals and glitches using a ``coherence test'' where the ``coherent'' signal hypothesis uses a template-based analysis assuming the data streams from multiple detectors contain a coherent gravitational wave signal while the ``incoherent'' glitch hypothesis treats each data stream independently.  The incoherent model uses the same template waveform as the signal model but optimizes its parameters independently for each detectors' data~\cite{Veitch:2009hd}.

Recently we proposed \BayesWave{}--a Bayesian algorithm to follow-up short duration ($\lesssim 1$ s) candidate gravitational wave transient events, separate signals from glitches, and provide robust signal characterization for arbitrary burst waveforms~\cite{Cornish:2014kda}.  \BayesWave{} uses a variable-dimension model for signals and/or glitches enabling the analyses to adapt the complexity of the waveform model to match what is present in the data instead of imposing a template waveform and searching for best fit parameters.
For a detection candidate \BayesWave{} computes the relative evidence of the event being produced by a GW signal, an instrument artifact, or statistical fluctuations of the detector's Gaussian noise.  
In the event that the candidate is of astrophysical origin, \BayesWave{} also produces posterior distributions for the source sky-location and orientation, accurate waveform reconstruction, and metrics to characterize the signal such as duration, bandwidth, signal energy, etc.   
In all instances, \BayesWave{} characterizes the instrument behavior including spectral estimation for the background Gaussian noise and glitch reconstructions which can then be used to feedback into the never-ending effort to improve the interferometers' performance.   
Analysis of the Gaussian component of the instrument noise is handled by \BayesWave{}'s sibling algorithm, \BayesLine{}~\cite{Littenberg:2014oda}.  During the first Advanced LIGO observing run \BayesWave{} is being utilized as a follow-up analysis to candidate and background events found by the coherent WaveBurst algorithm~\cite{Klimenko:2008fu}.

In this paper we will demonstrate \BayesWave{}'s potential by analyzing data from the sixth LIGO science run (S6) which took place from 2009-2010.  
Our results are achieved by analyzing data known to contain glitches which contributed to the long-tailed background distribution for the burst search, and by adding simulated gravitational wave signals to detector noise.
In addition to this study using archived data, we present an analytical framework for understanding the performance of the pipeline.
A companion paper uses \BayesWave{} and the flagship burst search algorithm, coherent WaveBurst~\cite{Klimenko:2008fu}, in an end-to-end demonstration of how burst detection efficiency is improved by the joint analysis~\cite{Kanner:2015}.

In section~\ref{sec:method} we briefly describe the \BayesWave{} algorithm, Bayesian model selection, and our model for the data.  
In section~\ref{sec:math} we go through a simple analytic calculation to give insight into how \BayesWave{} is able to distinguish signals and glitches, and use \BayesWave{}'s performance on simulated signals added to real data to support the analytic approximations.
Section~\ref{sec:background} uses the intuition built from the analytics to estimate background rates for glitches to be considered signals by \BayesWave{}, and connects the Bayes factor to false alarm rates for detections.
We summarize the work in section~\ref{sec:discussion}.  The appendix contains a more detailed derivation of the analytic approximation to the evidence.

\section{Method}\label{sec:method}
Searches for Burst signals have been based on frameworks that employ detection statistics to measure the likelihood that Gaussian noise could produce the data \citep{Chatterji:2004qg,Chatterji:2005,Chatterji:2006nh,Klimenko:2008fu,Sutton:2009gi}.
While stationary Gaussian noise is often a good description for LIGO/Virgo data, the approximation breaks down with much higher regularity than the arrival of detectable gravitational waves.
Any data analysis method must account for the possibility of non-stationary non-Gaussian noise. 
Most existing analysis strategies apply various selection cuts to separate glitches from astrophysical signals which are tuned by adjusting thresholds to minimize the estimated background rate of transient noise glitches~\cite{Abbott:2009zi,Abbott:2009up,Abadie:2012rq}. 

Bayesian hypothesis testing has been used in searches for GWs from a timing glitch in the Vela pulsar~\cite{Abadie:2010sf} using a damped sinusoid that abruptly starts at times associated with the pulsar timing glitch as the signal model.  A recently developed search pipeline~\cite{oLIB} uses excess power to identify interesting data segments and a matched-filtering follow-up with a sine-Gaussian template for signal characterization~\cite{Essick:2014wwa}.  The ``coherent vs. incoherent'' Bayes factor is used to distinguish between noise and signal~\cite{Veitch:2009hd}. 

\BayesWave{} employs a different approach by using a parameterized model for the LIGO/Virgo data, noise and signal included, and forward modeling, i.e. predicting, the detector output.
The data model has three distinct components:  A gravitational wave signal $h$ that is elliptically polarized and is coherent across the network of detectors;  glitches that are independent in each interferometer; and stationary Gaussian noise which is fully characterized by its power spectral density $S_n(f)$ as modeled by \BayesLine{}~\cite{Littenberg:2014oda}.  
At its core, \BayesWave{} is a Markov chain Monte Carlo (MCMC) algorithm~\cite{Gamerman:1997}.  
\BayesWave{} uses parallel tempering~\cite{Swendsen:1986} and thermodynamic integration~\cite{Goggans:2004} to compute the evidence for each model.  
The MCMC implementation and evidence calculation is described in detail in Refs.~\citep{Cornish:2014kda,Littenberg:2014oda}.  
For results in this work we utilize an adaptive temperature scheme as suggested in~\cite{Vousden:2015}.

Because we do not know \textit{a priori} the functional form of glitch or GW burst waveforms, our model for both must be flexible. 
We use a linear combination of Morlet-Gabor wavelets as our waveform model where the number of wavelets included in the linear combination, $N$, is itself a model parameter.
Each basis function (wavelet) is described by parameter vector $\intp\rightarrow \{f_0,t_0,A,Q,\varphi_0\}$ with components for central frequency $f_0$ ; central time $t_0$ ; amplitude $A$; quality factor $Q$;
and phase offset $\varphi_0$.  
A wavelet is expressed in the time domain as:
\begin{equation}
\Psi(t; A, f_0, Q, t_0, \phi_0) = A e^{-\Delta t^2 / \tau^2 } \cos (2 \pi f_0 \Delta t + \varphi_0 )
\end{equation}
where $\tau = Q/(2 \pi f_0)$ and $\Delta t = t - t_0$.  
\BayesWave{} uses a reversible jump Markov chain Monte Carlo~\cite{Green:2003} to marginalize over the number of wavelets needed for the model to be consistent with the data.

\subsection{Bayesian hypothesis testing or model selection}
The likelihood that hypothesis $\Hyp$, parameterized by $\params$, would have produced the data $d$ is calculated by
\begin{equation}\label{eq:evidence}
p(\data |\Hyp) = \int  p(\data|\params,\Hyp) p(\params|\Hyp) d\params. 
\end{equation}
Directly integrating Eq~\ref{eq:evidence} is seldom practical and a wide variety of alternative means for arriving at $p(\data|\Hyp)$ have been devised.
Our method of choice for computing the integral in Eq.~\ref{eq:evidence} is thermodynamic integration~\cite{Goggans:2004}.

Once the evidence has been computed it provides a relative measure of how well one hypothesis is supported by the data over another through the ``odds ratio''  
\begin{equation}
\Odd_{0,1}  \equiv \frac{p(\Hyp_0)}{p(\Hyp_1)}\frac{p(\data|\Hyp_0)}{p(\data|\Hyp_1)} = \frac{p(\Hyp_0)}{p(\Hyp_1)}\Bay_{0,1} 
\end{equation}
where $p(\Hyp)$ is the prior probability for the hypothesis and $\Bay_{0,1}$ is the likelihood ratio, or ``Bayes factor,'' for the two hypotheses.

\subsection{Modeling signals versus glitches}
Consider a GW network consisting of the two LIGO detectors in Hanford, WA (H) and Livingston, LA (L).
For a candidate event \BayesWave{} calculates the Bayesian evidence for each of three models: signal, glitch, or Gaussian noise.  
We can then use the Bayes factor between any two models to quantify the degree of supporting evidence for one model over the other.
Within each model the likelihood is computed by
\begin{equation}\label{eq:likelihood}
p(\data|\params,\Hyp) \propto \prod_I^{H,L}e^{-\frac{1}{2}(r^I(\params|\Hyp)|r^I(\params|\Hyp))}
\end{equation}
where $r^I$ is the residual of the data minus the signal or glitch model, $(a|b)\equiv \frac{2}{T}\int\frac{\tilde{a}^*(f)\tilde{b}(f) + \tilde{a}(f)\tilde{b}^*(f)}{S_n(f)}df$ is the noise weighted inner product, $S_n(f)$ is the noise power spectral density estimated from the data by \BayesLine{}, and $T$ is the duration of the data.  
This work will focus only on examples where we need to distinguish between the signal and glitch models.  We assume either will be preferred over the Gaussian noise model.

  \subsubsection{$\Hyp_0$:  The glitch model $(\Gli)$:}
  The data $d^I = n^I + g^I$ contain Gaussian noise $n$ and glitches $g$ independent in each detector $I$.
  The parameters $\params_\Gli \rightarrow [\intp^H \cup \intp^L]$ are comprised of independent sets of intrinsic parameters \[\intp^I\rightarrow[\intp_0 \cup \intp_1 \cup \cdots \cup \intp_{N^I}]\] which determine the shape of each wavelet.  The glitch model is computed for each detector as an independent linear combination of wavelets \[g(\intp^I,N^I) = \sum_i^{N^I} \tilde\Psi(f;\intp^I_i) \]
where $\tilde\Psi(f)$ is the Fourier transform of $\Psi(t)$, $N^I$ can take on any value between $[0,N_{\rm max}]$ with the caveat that \emph{at least one wavelet} must be used in the model for the whole network.  $N_{\rm max}$ is typically 20.  The glitch-model likelihood is computed using Eq.~\ref{eq:likelihood} with the residual $r^I(\params,\Gli)=d^I - g(\intp^I,N^I)$
  
  \subsubsection{$\Hyp_1$:  The signal model $(\Sig)$:}
  The data $d^I = n^I + h^I$ contain Gaussian noise $n$ and an elliptically polarized gravitational wave signal $h$ coherent across the network of detectors.
  The parameters $\params_\Sig \rightarrow [\intp^\oplus \cup \extp]$ are a common set of intrinsic parameters \[\intp^\oplus\rightarrow[\intp_0 \cup \intp_1 \cup \cdots \cup \intp_{N^\oplus}]\] referenced at the center of the Earth and four ``extrinsic parameters'' \[\extp\rightarrow[\theta,\phi,\psi,\epsilon]\] which define the sky-location $\theta,\phi$, the polarization angle $\psi$ and an ellipticity parameter $\epsilon$ relating the two gravitational wave polarizations $h_+$ and $h_\times$.  
The signal-model likelihood is computed using Eq.~\ref{eq:likelihood} with the residual \[r^I(\params,\Sig)=d^I - h^I(f;\intp^\oplus,N^\oplus,\extp)\]

The geocenter signal wavelets are projected onto the network using each detector's unique time delay operators $\Delta t(\theta,\phi)$,
and antenna beam pattern response functions $F^+(\theta, \phi, \psi)$, $F^\times(\theta, \phi, \psi)$~\cite{Anderson:2001}:
\begin{eqnarray}\label{eq:project}
h^I(f;\intp^\oplus,N^\oplus,\extp) &=& \left(F_I^+h_+ (f)+ F_I^\times h_\times(f)\right) e^{2\pi i f \Delta t_I} \nonumber \\
h_+ (f) &=&\sum_i^{N^\oplus}\tilde\Psi(f;\intp_i^\oplus) \nonumber \\
h_\times(f) &=& \epsilon\, h_+(f) e^{i\pi/2}.
\end{eqnarray}

\section{Distinguishing signals from glitches}\label{sec:math}
While \BayesWave{} uses a computationally expensive numerical integration to compute the evidence for each model, we will build intuition for how \BayesWave{} successfully distinguishes signals from glitches using the Laplace approximation to the evidence and several simplifying assumptions about the model and the data.
As our results will show, the simple analytic treatment derived here leads to useful approximations for when signals and glitches are distinguishable and in forecasting the most significant background event.  A more detailed derivation and discussion of the Laplace approximation to \BayesWave{}'s signal and glitch model evidence can be found in the appendix.
\subsection{Laplace-Fisher approximation to the evidence}
If an event has enough SNR to be a strong candidate for detection (${\rm SNR} \equiv \sqrt{(h|h)}\gtrsim  10$) the integrand of Equation (\ref{eq:evidence}) will be sharply peaked around the maximum a posteriori (MAP) parameter values of the model $\params_{\mathrm{MAP}}$. 
The evidence can be estimated as 
\begin{equation}
p(d|\Hyp) \simeq p(d|\params_{\rm MAP}, \Hyp)p(\params_{\rm MAP} | \Hyp) (2\pi)^{D/2} \sqrt{\det C}
\end{equation}
which is the product of the MAP likelihood $p(d|\params_{\rm MAP}, \Hyp)$, the prior $p(\params_{\rm MAP} | \Hyp)$ evaluated at the MAP parameters, and the determinant of the parameter covariance matrix $C$ which is a measure of the posterior volume.  $D$ is the dimension of the model.  The covariance matrix $C$ can be approximated by the inverse of the Fisher information matrix $\Gamma$, and we replace $\sqrt{\det C}$ with $1/\sqrt{\det \Gamma}$.  

The  $p(\params_{\rm MAP} | \Hyp) (2\pi)^{D/2} \sqrt{\det C}$ term is the ``Occam factor'' that penalizes the likelihood by the model's size.  
If two models achieve the same likelihood the Occam factor, and therefore the evidence, will be smaller for the model that requires more (constrained) parameters to achieve that fit.
Consider a simple model with a single parameter $x$ and uniform prior over an interval $V_x$.  
The covariance matrix is simply the variance of the likelihood $\sigma_x^2$.
In this case the Occam factor is proportional to $\sigma_x/V_x$ which leads to a simple, intuitive, interpretation:  \emph{The Occam factor is the fraction of the prior taken up by the posterior}.
We will return to this interpretation when predicting the most significant background event for \BayesWave{}.  

For the glitch or signal model, the expectation value for the intrinsic parameter log likelihood is proportional to~\cite{Rover:Thesis} 
\begin{equation}
\ln p(\intp_{\rm MAP} | \Hyp) \sim \frac{ {\rm SNR}^2}{2}+\frac{D}{2}.
\end{equation} 
For uniform priors $p(\intp_{\rm MAP} | \Hyp) = 1/V_{\intp}$ where $V_{\intp}$ is the volume of the intrinsic parameter space.  
\BayesWave{} uses uniform priors for all but the amplitude parameter, with $p(A)$ a function of the wavelet's SNR~\cite{Cornish:2014kda}. 
For simplicity we will neglect the parameter-dependence of the amplitude prior in favor of the simpler $1/V$ scaling.
A similar but more detailed derivation including the true amplitude used by \BayesWave{} can be found in the appendix.

The determinant of the intrinsic parameter Fisher matrix for a single wavelet is 
\begin{equation}
\det\Gamma_\intp = \frac{\pi^2 {\rm SNR}^{10}}{2 Q^2}.
\end{equation}
If we assume little overlap between wavelets in the parameter space the correlations between wavelet parameters are negligible and the Fisher matrix is block diagonal.  
The determinant for the full covariance matrix with $N$ wavelets is then~\cite{Cornish:2014kda} 
\begin{equation}
\sqrt{\det C_\intp} \approx \prod_n^N \frac{\sqrt{2}Q_n}{\pi{\rm SNR}_n^5}.
\end{equation}

Neglecting the extrinsic parameters for the signal model, and the \BayesLine{} parameters which are common to all models, the dimension $D=5N$ where $N$ is the number of wavelets used in the fit. 
To simplify the expression we define $\bar{Q}_n\equiv (2\pi)^{5/2}\frac{\sqrt{2}Q_n}{\pi}$ to absorb the $(2\pi)^{D/2}$ and additional factors of 2 and $\pi$.  Now the log evidence becomes
\begin{equation}\label{eq:logZ}
\log p(d|\Hyp) \simeq \frac{{\rm SNR}^2}{2} + \frac{5N}{2} - N\log(V_\intp) + \sum_n^N \frac{\bar{Q}_n}{{\rm SNR}_n^5}.
\end{equation}
From this expression we see that the Bayes factor for either the glitch or signal model versus the Gaussian noise model goes as $\mathcal O\left({\rm SNR}^2\right)$. 

For the glitch model, the prior and posterior volume terms are summed over the number of detectors (${\rm IFO}$) in the network.  The signal model, on the other hand, picks up an additional $D_\extp/2$ and Occam factor term $\log \sqrt{\det C_\extp}/V_\extp$ for the extrinsic parameters which govern the projection of the signal onto the network.  $D_\extp$ is the extrinsic parameter dimension, $C_\extp$ is the signal parameter covariance matrix, and $V_\extp$ is the extrinsic parameter prior volume.  Including these details into Eq.~\ref{eq:logZ} we find the log evidence for the glitch and signal models is
\begin{widetext}
\begin{eqnarray}\label{eq:logZsg}
\log p(d|\Gli) &\simeq& \frac{{\rm SNR^2_{\rm NET}}}{2} +  \sum_i^{\rm IFO}\frac{5N^\Gli_i}{2}-\sum_i^{\rm IFO}N^\Gli_i\log(V_\intp) + \sum_i^{\rm IFO}\sum_n^{N^\Gli_i} \frac{\bar{Q}_{i,n}^\Gli}{{\rm SNR}_{i,n}^5} \nonumber \\
\log p(d|\Sig) &\simeq& \frac{{\rm SNR^2_{\rm NET}}}{2} + \frac{5N^\Sig}{2} - N^\Sig\log(V_\intp) + \sum_n^{N^\Sig} \frac{\bar{Q}_{n}^\Sig}{{\rm SNR}_{{\rm NET},n}^5} + \left[ \frac{D_\extp}{2} + \log\frac{\sqrt{\det C_\extp}}{V_\extp}\right]
\end{eqnarray}
\end{widetext}
respectively, where ${\rm SNR}^2_{\rm NET} = \sum_i^{\rm IFO}{\rm SNR}_i^2$ and the extrinsic parameter dimension $D_\extp=4$ while the prior volume for extrinsic parameters is $4\pi^2$.  

\subsection{Two detector network}
Consider a fairly loud gravitational wave signal in the two detector LIGO network.  
The optimal extrinsic parameters for detection will result in similar signal strength in each of the interferometers such that ${\rm SNR}_{I,n}\approx{\rm SNR}_{{\rm NET},n}/\sqrt{2}$ where the index $n$ is for each wavelet and the index $I$ is for each detector.
For such events the glitch model will use similar wavelets as the signal model $Q^\Gli_{I,n} = Q^\Sig_n = Q$, but because it treats each detector independently, will need two copies--one for each interferometer $N^\Sig = N^\Gli_I = N$.  
One final simplifying assumption is that the signal to noise ratio of each wavelet is the same: ${\rm SNR}_{{\rm NET},n} \approx \overline{\rm SNR} = {\rm SNR}_{{\rm NET}}/\sqrt{N}$.

Substituting these simplifications into Eq.~\ref{eq:logZsg} we arrive at a simple expression for the log Bayes factor $\log\Bay_{\Sig,\Gli} = \log p(d|\Sig) - \log p(d|\Gli)$:
\begin{widetext}
\begin{equation}\label{eq:approx_bayes}
\log\Bay_{\Sig,\Gli} \simeq \frac{5N}{2} + N\log V_\intp + 5N\log(\overline{\rm SNR}) - \sum_n^N\log\bar{Q}_n + \left[2 + \log\frac{\sqrt{\det C_\extp}}{4\pi^2}\right]
\end{equation}
\end{widetext}
and immediately see that $\log\Bay_{\Sig,\Gli}\sim\mathcal O\left(N\log{\rm SNR}\right)$.  As a consequence, at fixed SNR, waveform morphologies that require more wavelets to reconstruct have a higher likelihood of being classified as signals.
This is an important departure from traditional SNR-based ranking statistics.  
The Bayes factor computed by \BayesWave{} is more sensitive to signal complexity than signal strength. 
Heuristically, the $\log\Bay_{\Sig,\Gli}$ naturally encodes how increasingly unlikely it is for the detectors to simultaneously and coherently produce glitches with non-trivial time-frequency structure.  This is a significant difference from existing Burst pipelines which put greater emphasis on signal strength in forming their detection statistic, and are thus hamstrung by the detectors' tendency to produce loud noise transients at a higher rate than the universe supplies gravitational wave signals.
We find this fundamental difference allows \BayesWave{} to assign detection candidates high confidence in data prone to loud glitches while existing pipelines have not.  

\subsection{Single wavelet examples in simulated noise}

To verify the predictions from the Laplace approximation we used \BayesWave{} to recover simulated sine Gaussian gravitational wave signals in Gaussian noise, drawing waveform parameters from our prior distributions:  $f\in[16,512]$ Hz, $t\in[-0.5,0.5]$ s, $Q\in[3,40]$, $\varphi\in[0,2\pi]$ rad, $\cos\theta\in[-1,1]$, $\phi\in[0,2\pi]$ rad, $\psi\in[0,\pi/2]$ rad, $\epsilon\in[-1,1]$, and amplitude drawn from the distribution described in the appendix and Ref.~\cite{Cornish:2014kda}.
For this study we analyze segments of LIGO data collected during the sixth science run which took place from 2009-2010 in which we have purposefully added GW signals.  The priors used for this analysis reflect what is being used for low-frequency triggers in the first advanced LIGO observing run (O1) during which \BayesWave{} relies on the coherent WaveBurst pipeline to provide the segments of data which warrant follow-up analysis (for details see Refs.~\cite{Klimenko:2008fu,Kanner:2015}). 

\BayesWave{} calculates Bayes factors for each combination of models along with an estimate of the error in that calculation, using thermodynamic integration.
We do not anticipate the agreement between numerical simulations and the analytic approximations to be perfect.  
Many of the approximations we have made along the way to arrive at Eqs.~\ref{eq:logZ} and~\ref{eq:approx_bayes} are known to be inadequate for the gravitational wave detection problem~\cite{Cornish:2007if}, particularly our use of the covariance matrix to estimate the posterior volume and, even more egregiously,  our use of the Fisher matrix as the inverse covariance matrix~\cite{Vallisneri:2007ev}.

Fisher matrix approximations are particularly bad for $\det C_\extp$.  The extrinsic parameter space is rife with degeneracies between parameters and non-Gaussian, multimodal likelihood distributions which often span the full extent of the prior range.  Fisher matrix arguments would predict a $\rm SNR^{-D_\extp}$ scaling for the determinant of $C_\extp$ which is much too strong for any burst source in a two detector network.  Using numerical experiments to get a rough understanding of the extrinsic parameter posterior volume, we find an $\rm SNR^{-\gamma}$ with $\gamma$ ranging from $\sim 1$ at low SNRs  to $\sim 2.5$ at $\rm SNR\sim100$ (See Fig.~\ref{fig:extrinsic_volume} in the appendix).

In Figure~\ref{fig:prediction} the left panel shows the glitch to noise (red) and signal to glitch (blue) log Bayes factors as a function of the simulated signals' SNR along with the Laplace approximation predictions.  The predicted scaling laws for $N=1$ signals  $\log\Bay_{[\Sig/\Gli],\mathcal N}\sim\mathcal O\left({\rm SNR}^2\right)$ and $\log\Bay_{\Sig,\Gli}\sim\mathcal O\left(\log{\rm SNR}\right)$ are generally obeyed by the numerical results.  The observed agreement reinforces the intuition developed from considering the analytic expressions, and we can be confident that the numerical integration is performing well.  The right panel demonstrates \BayesWave{}'s glitch rejection capabilities by comparing $\log\Bay_{\Sig,\Gli}$  for simulated sine-Gaussian glitches (gray crosses) and signals (blue circles).  The glitches were simulated by adding sine-Gaussians to each detector with parameters drawn independently from the prior.  Negative $\log\Bay_{\Sig,\Gli}$  corresponds to data with higher likelihood for the glitch model.

\begin{figure*}
\begin{center}
\mbox{
\includegraphics[width=\linewidth]{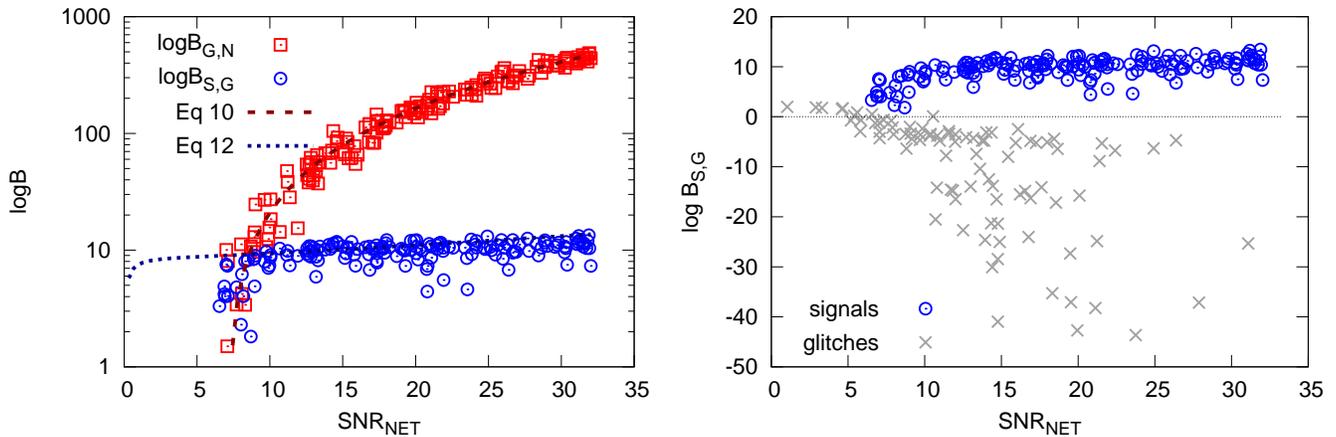} 
}
\caption{Left panel:  Comparing the numerical results for $\log\Bay_{\Gli,{\mathcal{N}}}$ (red squares) and $\log\Bay_{\Sig,\Gli}$ (blue circles) to the relevant analytic predictions in Eqs.~\ref{eq:logZ} and \ref{eq:approx_bayes} (dark red dashed and dark blue dotted lines, respectively) showing good agreement at high SNRs where the Laplace approximation is more valid. Right panel:  The  $\log\Bay_{\Sig,\Gli}$ results from the left panel (blue circles) with $\log\Bay_{\Sig,\Gli}$ from a set of glitches simulated by adding independent sine-Gaussian waveforms to each detector (gray crosses).  The log Bayes factor shows a clear separation between the simulated signals and glitches.  Cases with $\log\Bay_{\Sig,\Gli}<0$ correspond to the glitch model having a higher likelihood.}
\label{fig:prediction}
\end{center}
\end{figure*}

\subsection{Multiple wavelet examples in real noise}

Equation~\ref{eq:approx_bayes} predicts that the Bayes factor grows with SNR more rapidly for waveforms that have more time-frequency structure, thus requiring more wavelets to account for all of the excess power in the data.  
For astrophysical signals the number of wavelets necessary will not be known a priori, and furthermore will not be constant, depending on the SNR.
As the signal strength increases, more detailed structure in the waveforms will be detectable, and more wavelets will be favored by the model selection.
Through numerical experiments we find simple relationships for the number of basis functions and the average SNR per wavelet in terms of the true SNR:
\begin{eqnarray}
N\sim1+\beta{\rm SNR}\nonumber\\
\overline{SNR}\sim\alpha{\rm SNR}^a
\end{eqnarray}
where the coefficients $\beta,\alpha$ and index $a$ are different for different kinds of signals with $\{\beta{\sim}0,\alpha{\sim}1,a{\sim}1\}$ corresponding to sine Gaussian waveforms and $\beta$ and $\alpha$ increasing while $a$ decreases with increasing signal complexity (see Fig.~\ref{fig:prediction_laplace} in the appendix).

\begin{figure}
\begin{center}
\mbox{
\includegraphics[width=\linewidth]{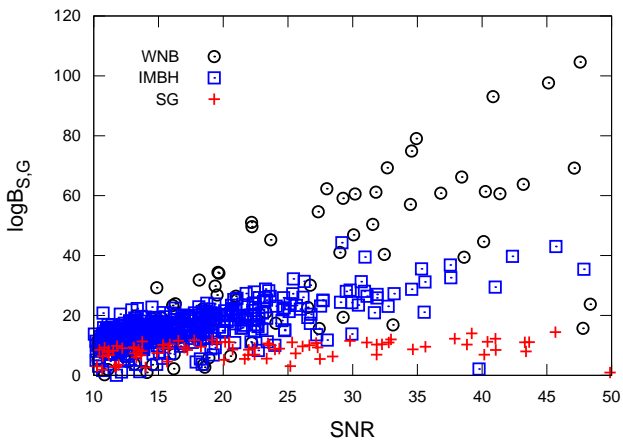} 
}
\caption{Bayes factors between the signal and glitch model grow more rapidly with SNR for more complicated waveform morphologies.  Sine Gaussian injections (red) only require a single wavelet and thus never achieve particularly high $\log\Bay_{\Sig,\Gli}$.  Intermediate mass black hole (IMBH) mergers need $N \lesssim 10$ resulting in stronger separation between models (blue), while the white noise burst waveforms (black) have rich time frequency structure, often saturating the prior on $N$ and show the strongest SNR dependence (highest $\beta$).}
\label{fig:evidence}
\end{center}
\end{figure}

To demonstrate this important feature of \BayesWave{} we add simulated gravitational wave signals from different waveform families into real detector data. 
Figure~\ref{fig:evidence} shows $\log\Bay_{\Sig,\Gli}$ as a function of SNR for the different simulations.  Red points are sine Gaussian waveforms, blue points correspond to signals from the merger of two 50 M$_\odot$ black holes modeled using non-spinning Effective One Body (EOB) waveforms~\cite{Buonanno:2007pf}, and the black points are results from ``white noise bursts''--unpolarized, band-limited, white noise signals used to test LIGO/Virgo burst detection pipelines.  We can empirically determine that $\beta$ is larger for more complicated signal morphologies.  Results agree well enough with the analytic predictions that the insight gained in the analytic study is applicable, but the Laplace approximation is clearly no substitute for the numerical integration.  The large scatter in Bayes factors is due to failings in the Laplace approximation, signals that violate our assumption about roughly equal SNR in each detector, and segments of data that contain both signals and glitches.

It is important to note that the high degree of scatter in the white noise burst results is also to be expected because these signals are unpolarized, while \BayesWave{} assumes  $h_+$ and $h_\times$ are related by Eq.~\ref{eq:project}.
In a two detector network we generally cannot reliably measure the GW polarizations independently.  Introducing the additional degrees of freedom to independently solve for $h_+$ and $h_\times$ will hinder our ability to reject glitches because the number of signal model and glitch model parameters will be comparable for a wider variety of waveform morphologies.  
While there is no reason to expect a priori that GW bursts will be elliptically polarized, selection effects by the detection pipelines which identify segments of data for \BayesWave{} to follow-up in a real analysis, and the similar orientation of the LIGO detectors, favor signals which are well approximated by a single polarization (causing many of the degeneracies between extrinsic parameters discussed in the previous section).
This assumption will need to be relaxed when more detectors are added to the network, and in future studies we will investigate strategies for optimizing \BayesWave{}'s performance on unpolarized detection candidates even in the two detector case.

\section{Background estimation}\label{sec:background}
We have shown that \BayesWave{} predictably favors the signal model over the glitch model for simulated GW events, i.e. \BayesWave{} is robust against false dismissal of gravitational wave signals.
This is only half of the battle:  Any useful data analysis procedure must also be robust against false alarms, i.e. misidentifying noise events as being astrophysical signals, and be able to assign significance to a detection.  While the right panel of Figure~\ref{fig:prediction} demonstrates how \BayesWave{} can reject glitches in the trivial case of random sine-Gaussian waveforms, how it will fair against real glitches, and how to assign significance to candidate events, requires more careful attention. 

To understand \BayesWave{}'s glitch rejection capabilities, imagine that a glitch waveform in LIGO Hanford ($H$) is well represented by a linear combination of wavelets with parameters $\intp^H$ and a coincident (i.e. within the light travel time between detectors) collection of wavelets is found in LIGO Livingston ($L$).  
If the signal is astrophysical in nature, the waveform in $L$ must have parameters $\intp^L$ that are consistent with $\intp^H$, within the measurement uncertainties $\sqrt{\det C_{\intp^L}}$ up to the appropriate time, phase, and amplitude shifts due to the geometry of the detector locations and orientations.  On the other hand, if the data represent coincident glitches, then \textit{a priori} there is no reason for the glitch in Livingston to match the parameters in Hanford.
Instead, the wavelet in Livingston is chosen at random.
One can consider glitches to be random draws from $\intp$ space and false alarms (glitches that appear as signals) are draws that overlap within the size $\sqrt{\det C_{\intp^L}}$.
If the posteriors do not overlap the data is not consistent with the signal model, i.e. the signal model likelihood will be lower than the glitch model likelihood, and the Bayes Factor will favor the glitch model (c.f. Figure~\ref{fig:prediction}).

We can use the same logic to estimate the background rate of glitches that are consistent with the signal model.
Assume, in a given two detector data set, there are $\mathcal{N}_{\rm gl}$ coincident glitches. 
If we assume our signal/glitch model can achieve a perfect match to glitches in the data, the recovered SNR of the signal and the glitch model will be equal when the glitches overlap in the $\intp$ space and the Bayes factor will be again well approximated by Eq.~\ref{eq:approx_bayes}.

Recall the Occam factor is interpreted as the fraction of the prior covered by the posterior $\sim \sqrt{\det C}/V$, i.e. the Occam factor is the size of the ``target'' the second glitch must hit to be misidentified as a signal.  Put another way, a glitch has probability $\sim \sqrt{\det C}/V$ to be consistent with the signal model.  Therefore finding a background event with a Bayes factor consistent with Eq~\ref{eq:approx_bayes} will require analysis of something like $ \left(V_\intp\right)^N/\sqrt{\det C_\intp}$ coincident glitches.  
In our application the Occam factor thus takes on an additional interpretation as the expected number of trials (coincident noise transients) needed for two random glitches to have sufficient overlap in parameter space to look like a signal.  

We can loosely turn this into an argument for the maximum Bayes factor--the one that occurs \emph{only once} in a span of LIGO data--as having an Occam factor of $N_{\rm gl}$, i.e. the maximum Bayes factor for a background noise event is
\begin{equation}\label{eq:loudest}
\langle\Bay_{\Sig,\Gli}\rangle_{\rm background} \lesssim N_{\rm gl}.
\end{equation}

This limit is not robust.  The loudest noise event is obviously in the extreme tail of the background distribution and will therefore fluctuate wildly for different realizations of the data. Nor is this a statement about the population of glitches beyond the assumption that the parameters $\intp$ are chosen at random for glitches in each detector. 
It is also important to point out this is may be a conservative estimate.  
Most glitches are at low SNR in any realistic glitch population, and so low values of the Occam Factor will likely be much more common than high values.

We use our estimate of the most significant background event to approximate the false alarm rate.  
To do so we need to know the rate of coincident glitches, $R_{\rm gl}$, which is a carefully studied quantity within LIGO.  
The single detector glitch rate was known during S6 to typically have values between 1 and 0.1 Hz~\cite{Aasi:2014mqd}.  
The light travel time between LIGO detectors is 10 ms, leading to a coincident glitch rate of
$R_{\rm gl} \sim 1\ \rm{Hz} \times 1\  \rm{Hz} \times 0.01 \ \rm{s} = 0.01\ \rm{Hz}$.

False alarm rates are estimated by analyzing time-shifted data, or ``time slides.''  
If the data from one detector is shifted by more than the light travel time to another detector, there will be no coincident gravitational wave signals.
Because the rate of glitches completely dominates the rate of GW signals, analyzing time-shifted data all but guarantees that any coincidences are due to noise artifacts.

Consider the last quarter of LIGO's sixth science run (S6D) which lasted for $\sim50$ days.   
A so-called ``three sigma'' detection requires an event more significant than any background coincidences found in $\sim 300$ time slides.
The background estimate from 300 time slides corresponds to 40 years of data, and $N_{\rm gl} \sim 1 \times 10^7$.  
Equation \ref{eq:loudest} predicts that events with $\ln \Bay_{\Sig,\Gli} \gtrsim 16$ would be detected with better than three-sigma confidence.  

\begin{figure}
\begin{center}
\mbox{
\includegraphics[width=\linewidth]{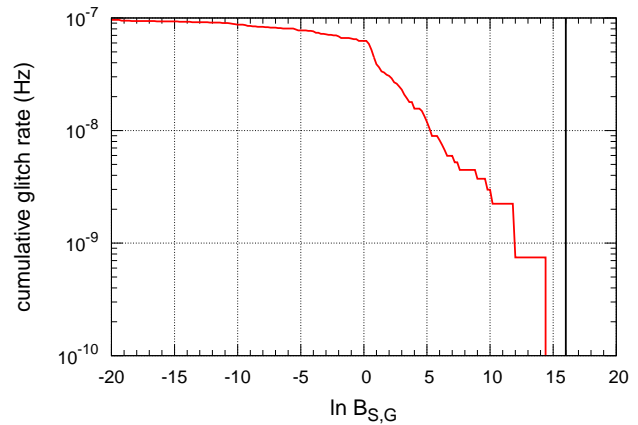} 
}
\caption{The cumulative rate of glitches as a function of Bayes factor from time-slide studies using \BayesWave{} and the S6D data set.
The black, vertical line shows the expected value for the most significant event using the limit in Equation \ref{eq:loudest}. 
Because this represents 300 time-slides of the data set, we see that the sine-gaussian injections above network SNR 10 are detected with marginal significance, whereas many WNB signals above network SNR 25 were ``gold-plated'' detections.  
Our findings are in excellent agreement with the analytic approximations in this work.}
\label{fig:cwb_bkg}
\end{center}
\end{figure}

To test this prediction we compute the Bayes factors for the coincident events in time slides of the S6D data found by the \texttt{coherent WaveBurst} algorithm~\cite{Klimenko:2008fu}.
Figure~\ref{fig:cwb_bkg} shows the cumulative glitch rate as a function of $\ln \Bay_{\Sig,\Gli}$ i.e. the y-axis is the rate at which coincident glitches were found with Bayes factors greater than the corresponding value on the x-axis.
The distribution steeply decreases with increasing Bayes factor, and does not show evidence of leveling-off with a broad ``tail'' in the background that has limited previous searches.   See Ref.~\cite{Kanner:2015} for a detailed study of how \BayesWave{} can improve detection confidence of existing burst searches.
Furthermore, the distribution ends at $\ln \Bay_{\Sig,\Gli}\sim15$ which is consistent with our analytic prediction for the background.
Ultimately we should be able to turn arguments about the expected background rate into a prior odds ratio between the glitch and signal model.  
For the immediate future we elect to take a more conservative approach and continue using background studies to estimate the false alarm rate and therefore the detection significance.
There is no guarantee that the non-Gaussian noise in future GW data will bear any resemblance to what was found during S6.

Comparing our inferred background rate to Figure~\ref{fig:evidence} we find that sine-Gaussian waveforms in a two detector network will be detected at false alarm rates that suggest marginal significance at any reasonable SNR, similar to performance seen in past burst searches. 
However, unlike previous burst searches, we find that IMBH and white noise bursts are detectable with very high significance.  Figure~\ref{fig:prediction_laplace} in the Appendix shows the number of wavelets used to recover each waveform morphology as a function of injected SNR and provides supporting evidence that waveforms that require more wavelets typically provide higher Bayes factors.

What is required for a high confidence, or ``five-sigma,'' detection?
For this case, we seek a p-value of less than $3 \times 10^{-7}$,
and so demand our event be louder than the loudest event in
$3 \times 10^6$ time slides.  For S6D this leads to $N_{\rm gl} \sim 10^{11}$,
and an expected loudest event $\langle \ln\Bay_{\Sig,\Gli}\rangle_{\rm background} \sim 25$.
We have already seen that single wavelet events can not reach this
level at any reasonable SNR but applying the scaling law
in Equation~\ref{eq:approx_bayes}, we find that such a ``gold-plated'' detection
could be achieved at reasonable SNR with as few as two or three wavelets.
For example, the IMBH and white noise burst signals in Figure~\ref{fig:evidence} added to the same data we used to estimate the background by far exceed the Bayes factor which corresponds to a false alarm probability of $\sim3\times10^{-7}$.
This is an important feature of the \BayesWave{} pipeline:
\emph{Gold-plated detections of short-duration signals are possible even in the presence of a significant glitch population.}

\begin{figure*}
\begin{center}
\mbox{
\includegraphics[width=\linewidth]{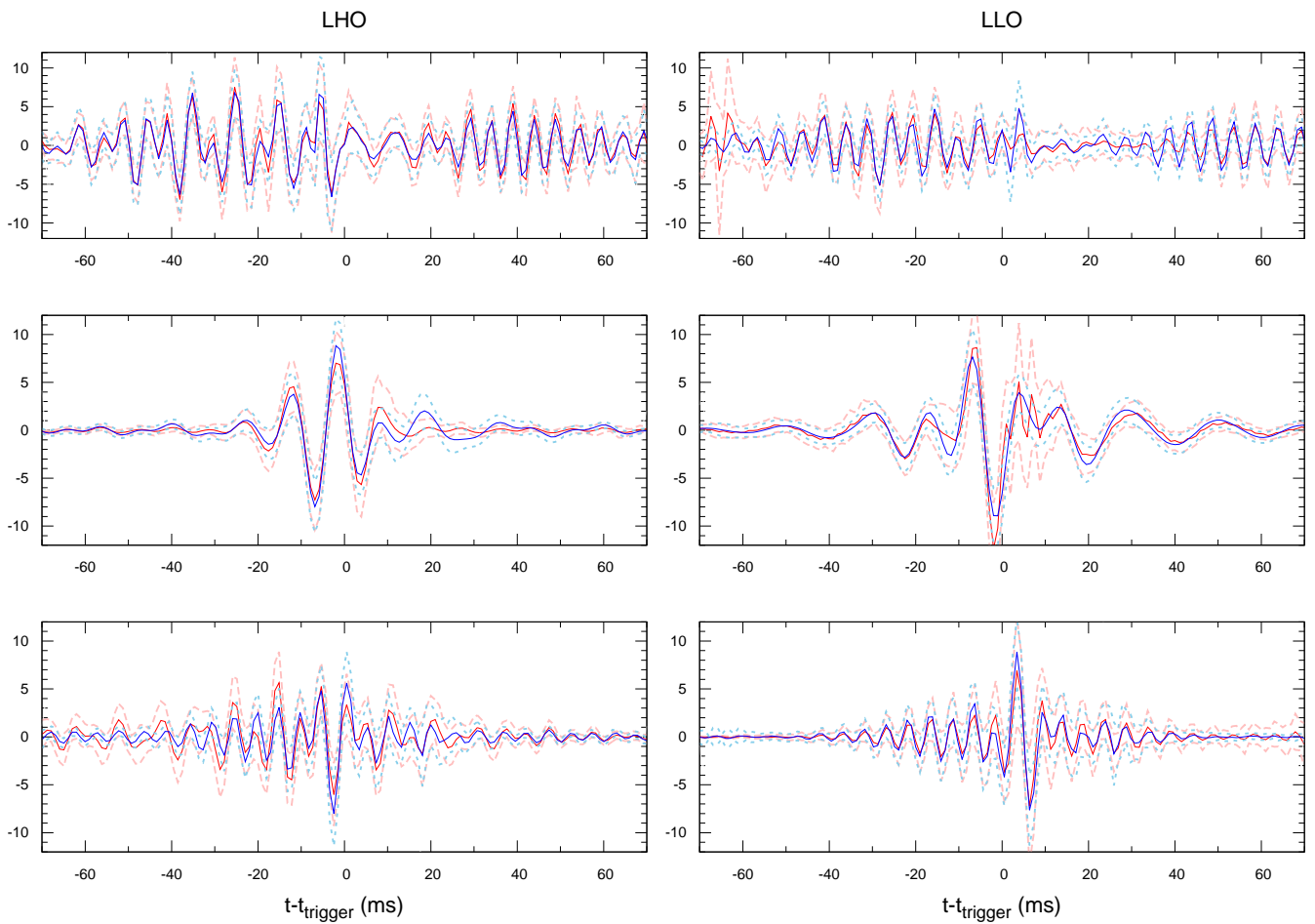} 
}
\caption{Reconstructed whitened, time-domain signal- and glitch-model waveforms for S6D background events.  Solid (red/blue) line is the median (glitch/signal) waveform.  Dashed lines of corresponding color show the $2\sigma$ errors on the reconstructed waveforms.  Each row shows one of the three most significant background events.  Left column is the waveform in Hanford.  Right column is Livingston.  From top to bottom $\ln\Bay_{\Sig,\Gli}$ was $[15, 12, 12]$.  The overlap between the glitch model and the signal model was $[91\%, 93\%,\ {\rm and\ } 86\%]$, respectively.}
\label{fig:cwb_alarms}
\end{center}
\end{figure*}

\section{Discussion}\label{sec:discussion}
In this paper we have demonstrated \BayesWave{}'s utility as a follow-up analysis for GW burst searches. 
By analyzing data from the sixth LIGO science run (S6) which took place from 2009-2010 we have shown that high confidence detections are achievable using \BayesWave{} as a follow-up analysis despite the high rate of noise transients in the data.
When used to follow-up short-duration gravitational wave triggers, \BayesWave{} has been shown to significantly reduce the rate of false-alarms while remaining sensitive to a wide range of signals~\cite{Kanner:2015}.  
For insight into how \BayesWave{} takes advantage of Bayesian model selection to separate signals and glitches we presented an analytical framework and found simple expressions which provide approximations to our full numerical analysis on real data.  The results show that \BayesWave{} has several novel features, when
compared with other Burst pipelines:
\begin{itemize}
\item The detection statistic directly compares the evidence for an astrophysical signal with a glitch model, as opposed to calculating a likelihood derived from Gaussian noise.
\item \BayesWave{} places emphasis on the time-frequency complexity and network coherence of an event, rather than just its strength, to distinguish signals from glitches
\item The background distribution shows no evidence of ``tails'' at high values.
\end{itemize}

In order to emphasize the importance of including the glitch model in a statistical framework, best fit waveforms for the signal and glitch models for 
the most ``signal-like'' background events in S6D are shown in Figure \ref{fig:cwb_alarms}.  For these examples of real glitches, the 
signal and glitch model are shown to very nearly agree.  Because glitches can be so successful in imitating real gravitational wave signals, 
pipelines which attempt to reject these events with tunings and cuts face a major challenge.  Instead, \BayesWave{} attempts to accurately 
assess the probability of such coincident glitches arising from chance.  This approach places a lower weight to events with simple time-frequency 
structure that could plausibly arise simultaneously in two or more instruments, regardless of their SNR.

The detection statistic described in this work, $\Bay_{\Sig,\Gli}$, represents the likelihood ratio
for two competing models: the data contains a glitch, or the data contains an astrophysical signal.
The purist may object to this application of the Bayes factor, instead favoring the Bayesian odds ratio between the signal and glitch model.
The prior odds ratio between these models is the ratio of the expected coincident and coherent glitch rate to the expected rate of GW signals.  While the rate of GW signals is unknown, we 
have shown that the measured background distribution is consistent with our analytic predictions using the LIGO 
glitch rate.  This consistency suggests that the \BayesWave{} model is a good fit to actual LIGO data and Bayes factors calculated by \BayesWave{} will serve as a robust means for correctly identifying signals and glitches.    
In principle, glitches with non-flat distributions in f and Q, especially if similarly distributed
in multiple detectors, could invalidate this agreement.  Should that be the case, the posterior distribution of background events can easily be folded in to our analysis as a prior on the glitch model.
Because the glitch population in earlier LIGO data will likely differ from that of the advanced detectors, we will continue to rely on the brute-force approach of using time slides to estimate the significance of a candidate event and use what is learned to further improve our priors for subsequently collected data.


As the capabilities of ground based detectors continues to improve so too must our analysis.  
The work presented here represents a snap shot of \BayesWave{}'s capabilities as the algorithm continues to advance.
Further development is underway to relax the requirement of elliptical polarization for the signal model (improving the detection efficiency for unpolarized signals) and to account for glitches and signals appearing in the same segment of data (reducing false dismissals due to near-coincidence with glitches).  
Nonetheless, based on the thorough performance studies in real LIGO data reported in this work we conclude that \BayesWave{} is prepared to decisively aid in the detection and characterization of GW bursts in the advanced detector era. 

\section{Acknowledgments}
We acknowledge numerous discussions with Reed Essick, Erik Katsavounidis, and Salvatore Vitale which helped motivate the detailed analytic derivation of the evidence in the Appendix; James Clark for thorough comments and suggestions on an earlier draft of this paper; Sergey Klimenko and Francesco Salemi for help finding and understanding the cWB output files; and  Kent Blackburn, Vicky Kalogera, Tjonnie Li, Patricia Schmidt, and Michele Vallisneri for helpful conversations about this work.  TBL acknowledges the support of NSF LIGO grant, award PHY-1307020.  LIGO was constructed by the California Institute of Technology and Massachusetts Institute of Technology with funding from the National Science Foundation and operates under cooperative agreement  PHY-0757058.  This paper carries LIGO Document Number LIGO-P1500083.

\appendix*
\section{Bayes Factors}

The Laplace approximation for the evidence is given by
\begin{equation}
Z = p(\data\vert \params_{\rm MAP},\Hyp) p( \params_{\rm MAP}|\Hyp) \, (2 \pi)^{D/2} \sqrt{\det C} 
\end{equation}
where $\params_{\rm MAP}$ are the maximum a posteriori parameters, $D$ is the model dimensions, and $C$ is the parameter covariance
matrix, which we can estimate from the inverse of the Fisher information matrix $\Gamma$. If the prior is uniform for all parameters, the prior density is equal to the
inverse of the prior volume: $p( \params_{\rm MAP}) = 1/V_{\rm prior}$. We recognize the collection of terms $ (2 \pi)^{D/2} \sqrt{\det C} $ as the posterior
volume. Thus, for uniform priors, the evidence is given by the product of the MAP likelihood times the ratio of the posterior to prior volume, which is referred to as
the Occam penalty. In the case of \BayesWave{} the priors on most parameters are flat, with the important exception of the amplitude or the signal-to-noise
ratio (which depends of the amplitude, quality factor, central frequency and noise spectral density).

Dropping terms down by factors of $e^{-Q^2}$ relative to leading order, the Fisher
matrix for a single wavelet using the parameters $\{t_0, f_0, Q, \ln A,  \phi_0\}$ is given by
\begin{equation}
\Gamma = {\rm SNR}^2 \left( \begin{array}{ccccc}
\frac{4 \pi^2 f_0^2 (1+Q^2)}{Q^2} & 0 & 0 & 0  & -2\pi f_0\\
0 & \frac{3 +Q^2}{4f_0^2} & -\frac{3}{4 Q f_0} & -\frac{1}{2 f_0} & 0 \\
0 &  -\frac{3}{4 Q f_0}  &  \frac{3}{4 Q^2}  & \frac{1}{2 Q} & 0 \\
0  &  -\frac{1}{2 f_0} & \frac{1}{2 Q}  & 1 &  0 \\
-2\pi f_0 & 0  & 0 & 0 & 1
\end{array} \right)
\end{equation}
The determinant of $\Gamma$ is
\begin{equation}
\det \Gamma = \frac{1}{\det C} =  \frac{\pi^2 \, {\rm SNR}^{10}}{2\, Q^2} \, .
\end{equation}

The expectation value of the MAP log likelihood is given by (see page 31 of Ref.~\cite{Rover:Thesis} and references therein)
\begin{equation}
{\rm E}[\ln p(\data\vert \params_{\rm MAP})] = {\rm const.} + \frac{D}{2}
\end{equation}
The constant is independent of the signal model. The $D/2$ term comes from more complicated models being able to better fit features in the Gaussian
noise.

Each wavelet is described by 5 parameters, and has 
\begin{equation}
\sqrt{\det C_{\intp_i}} = \frac{\sqrt{2} Q_i}{\pi\, {\rm SNR}_i^{5}}
\end{equation}
where ${\rm SNR_i}$ is the signal-to-noise ratio for wavelet $i$.  Assuming that the $N$ wavelets used in the reconstruction have little overlap with each other, 
the total posterior volume for the wavelet model is
\begin{equation}
\sqrt{\det C} = \prod_{i=1}^N \frac{\sqrt{2} Q_i}{\pi\, {\rm SNR}_i^{5}}
\end{equation}

\BayesWave{} has a non-trivial amplitude prior which needs to be taken into account. One choice would be a uniform in volume prior on the source
distribution, which is equivalent to a prior on the distance $D$ that scales as $p(D) ~ D^2$. Since amplitude and distance are inversely related, we have
$D^2 dD\sim A^{-4} dA\sim A^{-3} d \ln A$. Here we have made the change of variables to $\ln A$ since this is the parameter used to compute the Fisher
matrix. This prior is improper, and to make it proper a minimum amplitude cut-off $A_*$ (maximum distance) has to be introduced. The properly normalized
uniform-in-volume prior is 
\begin{equation}
p(\ln A) = 3 \left( \frac{A_*}{A} \right)^3 \, .
\end{equation}
An alternative approach, used by \BayesWave{} in this work, is to adopt different physically motivated priors on the
signals and glitches that are given as functions of the SNR. For glitches the SNR is given by
\begin{equation}
{\rm SNR} \simeq \frac{A \sqrt{Q}}{\sqrt{2 \sqrt{2 \pi} f_0 S_n(f_0)}} \, ,
\end{equation}
while for signals the SNR is given by the same expression, but with the individual detector noise spectral density replaced by 
the network average
\begin{equation}
\bar{S}_n(f_0)  = \left( \sum_i  \frac{F_{+,i}^2 + \epsilon^2 F_{\times,i}^2}{S_{n,i}(f_0)} \right)^{-1} \, .
\end{equation}
Thus the signal-model SNR depends on $A,Q,f_0,\extp$. The priors on the signal model and glitch model are given in terms of a prior
on the SNR, $p({\rm SNR})$. Making the change of variables from SNR to $\ln A$ yields
\begin{equation}
p(\ln A|\Gli) =  \left(\frac{\rm SNR}{{\rm SNR}_*}\right)^2 e^{-{\rm SNR}/{\rm SNR}_*}
\end{equation}
for the glitch model and
\begin{equation}
p(\ln A|\Sig) = \frac{3}{4}\left(\frac{\rm SNR}{{\rm SNR}_*}\right)^2 \frac{1}{ (1+ {\rm SNR}/4{\rm SNR}_*)^5}
\end{equation}
for the signal model.

If we further assume little correlation between the wavelet model and the Gaussian noise model, then 
the expected value for the log Bayes Factor between the glitch plus noise model and the noise model in a single detector is
\begin{widetext}
\begin{equation}
\ln \Bay_{\Gli,\Noi} =\frac{ {\rm SNR}^2}{2} +  \frac{5 N_\Gli}{2}\left(1  +  \ln(2\pi) \right)+\sum_{n =1}^{N_\Gli} \ln\left( \frac{\sqrt{2} Q_n}{\pi\, {\rm SNR}_n^{5}}\right)+\ln  p( \intp_{\rm MAP}|\Gli) \, .
\end{equation}
\end{widetext}
Here ${\rm SNR}^2$ is the signal-to-noise ratio of the signal or glitch in that detector. Later when considering a network of detectors the ${\rm SNR}^2$ will refer to the
network signal-to-noise ratio of the signal.

If the wavelet model in one detector uses $N$ wavelets, and assuming little overlap between wavelets, then
\begin{equation}
{\rm SNR}^2 = \sum_{n=1}^N  {\rm SNR}_n^{2} = N \, \overline{{\rm SNR}}^2
\end{equation}
Based on simulations, we find that the average number of wavelets used by \BayesWave{} increases linearly with the total ${\rm SNR}$, and that the average per wavelet $\overline{\rm SNR}$ increases with the total ${\rm SNR}$
as a waveform-dependent power law. Writing $N = 1+\beta{\rm SNR}$ and $\overline{{\rm SNR}}= \alpha\, {\rm SNR}^a$, we find that the values of $\beta$, $\alpha$, and $a$ depend on the waveform morphology, with
$\alpha$ and $\beta$ increasing, and $a$ decreasing, as the time-frequency structure of the waveform becomes more complicated. In the case of a constrained model using a fixed number of wavelets the average SNR per wavelet always increases linearly with the total SNR, though with a proportionality less than one for anything other than sine-gaussians.  

\begin{figure*}
\begin{center}
\mbox{
\includegraphics[width=\linewidth]{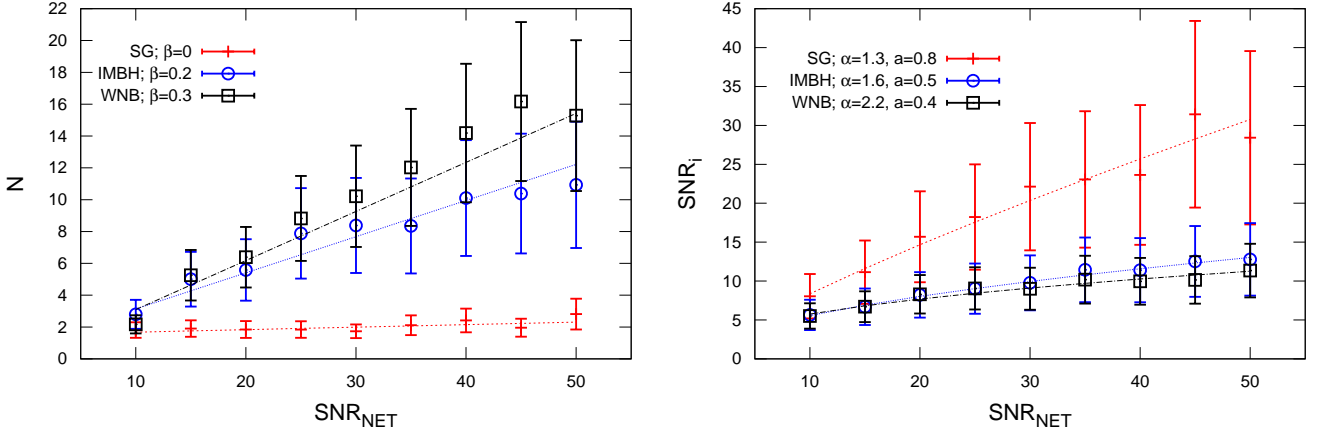} 
}
\caption{Number of wavelets (left) and SNR per wavelet (right) found for  sine Gaussian (red) intermediate mass binary black hole (blue) and white noise burst (black) waveforms in simulated Gaussian noise at full design sensitivity for a single Advanced LIGO detector averaged over several Gaussian noise realizations.  Error bars represent one standard deviation of the mean.  Over-plotted are SNR scalings found in the text using the best fit values $\beta$, $\alpha$, and $a$. }
\label{fig:prediction_laplace}
\end{center}
\end{figure*}

Fig.~\ref{fig:prediction_laplace} shows the average number of wavelets (left panel) and SNR per wavelet (right panel) as a function of SNR for the three different waveform morphologies studied in this paper--sine Gaussians, binary black hole mergers, and white noise bursts--added to simulated Gaussian noise from a single detector at Advanced LIGO sensitivity. Each simulation was repeated for several Gaussian noise realizations.  Plotted are the average and one standard deviations of the mean, plus lines that show the scaling relations using the best fit values for $\left\{\alpha,\beta,a\right\}$.

Starting with a simplified model of to aligned collocated detectors, the signal model does not
need any extrinsic parameters and the log Bayes factors are
\begin{widetext}
\begin{equation}
\ln \Bay_{\Sig,\Gli} = \left(\frac{5N_\Sig}{2}  - \frac{5N_\Gli}{2} \right) (1+\ln(2\pi))+\sum_{n =1}^{N_\Sig} \ln\left( \frac{\sqrt{2} Q_n}{\pi\, {\rm SNR}_n^{5}}\right)
- \sum_{n =1}^{N_\Gli} \ln\left( \frac{\sqrt{2} Q_n}{\pi\, {\rm SNR}_n^{5}}\right)+ \ln  p( \intp_{\rm MAP}|\Sig) -  \ln  p( \intp_{\rm MAP}|\Gli).
\end{equation}
\end{widetext}
If we assume that all the wavelets have the same quality factor $Q$ and individual signal-to-noise ratios ${\rm SNR}_i$, then the individual SNRs in each detector are ${\rm SNR}/\sqrt{2}$.  Thus for the glitch model ${\rm SNR}^2_i = \alpha\, {\rm SNR}/\sqrt{2}$, while for the signal model ${\rm SNR}_i = \alpha{\rm SNR}^a$.
Then the Bayes Factor between the signal and glitch models for two collocated detectors is   
\begin{widetext}
\begin{equation}
\ln \Bay_{\Sig,\Gli} = \frac{5 N_\Sig}{2} -  \frac{5 N_\Gli}{2} + N_\Sig \ln\left( \frac{4 \pi^{1/2} Q}{ {\rm TF}\, \Delta Q (\alpha {\rm SNR}^a)^{5}}\right)  - N_\Gli \ln\left( \frac{4 \pi^{1/2} Q}{ {\rm TF}\,  \Delta Q  (\alpha {\rm SNR}^a/\sqrt{2}))^{5}}\right)
+N_\Sig \ln p(\ln A|\Sig) -  N_\Gli \ln p(\ln A|\Gli)
\end{equation}
\end{widetext}
Here ${\rm TF}$ is the time frequency volume, and $\Delta Q$ is the prior range for $Q$. Note that the contributions from the amplitude prior introduce important SNR scalings
into the Bayes factor. For a single sine-Gaussian model the posterior volume terms introduce a $\ln({\rm SNR})$ scaling to the log Bayes factor.  For \BayesWave{} the scaling with SNR is far more complicated. On complex waveforms both $N_\Sig$ and $N_\Gli$ scale linearly
with the SNR, so the posterior volume introduces terms that scale as ${\rm SNR} \ln({\rm SNR})$. The amplitude prior for the signal model introduces a similar ${\rm SNR} \ln({\rm SNR})$
scaling, along with a more complicated scaling of the form ${\rm SNR} \ln(1+ {\rm SNR}/4{\rm SNR}_*)$. The amplitude prior for the glitch model introduces ${\rm SNR} \ln({\rm SNR})$ terms,
in addition to a term that scales as ${\rm SNR}^2$, though this term does not start to dominate until very high SNRs  (${\rm SNR} > 50$ for typical choices of ${\rm SNR}_*$). In the
fixed dimension case the \BayesWave{} scaling is dominantly of the form $\ln({\rm SNR})$ for moderate SNRs. At very high SNRs the quadratic dependence of the full \BayesWave{}
scaling is replaced by a linear scaling in SNR.

There are several assumptions that went into the derivation of the signal-to-glitch Bayes Factor for \BayesWave{} that are rather crude. The worst approximations are that the
wavelets used in the reconstruction all have roughly the same $Q$ and signal-to-noise ratio. While on average the scaling ${\rm SNR}_i = \alpha{\rm SNR}^a$ is quite robust,
the ${\rm SNR}_i$ for individual wavelets never go much below the value set by the peak of the SNR prior, ${\rm SNR}_*$ so that ${\rm SNR}_i \gtrsim {\rm SNR}_*$.
This means that the linear scaling typically only holds for network SNRs greater than around 10 or 12.
Rather than assuming the same quality factor for each wavelet we could use the average value. For $Q$ distributed uniformly in the range $Q\in [Q_1,Q_2]$ we have
\begin{equation}
{\rm E}\left[\ln \left(\frac{\Delta Q}{Q}\right)\right]= 1+\frac{Q_2\ln(\Delta Q/Q_2)-Q_1\ln(\Delta Q/Q_1)}{\Delta Q} \, ,
\end{equation}
and
\begin{equation}
{\rm Var}\left[\ln \left(\frac{\Delta Q}{Q}\right)\right]= \frac{Q_1Q_2(\ln(\Delta Q/Q_2)-\ln(\Delta Q/Q_1))^2}{\Delta Q^2}  - 1\, ,
\end{equation}

\subsection{Extrinsic Parameters}

Implicit in the preceding derivation was the assumption that the overlap of any two wavelets $(\Psi_i \vert \Psi_j){\sim}0$ and, as a consequence, the parameter correlation matrix for the wavelets was block-diagonal. This assumption is reasonable since each wavelet collects the power in a certain
time-frequency volume disfavoring highly overlapping wavelets.  For mis-aligned detectors we can write
the parameter correlation matrix for the signal model in block-form as
\begin{equation}
C = \left( \begin{array}{cc}
C_\intp  & C_X  \\
C_X^T & C_\extp \end{array} \right)
\end{equation}
where $C_\intp$ is the $5 N_\Sig \times 5 N_\Sig$ correlation matrix for the intrinsic wavelet parameters, $C_\extp$ is the $4 \times 4$ correlation matrix
for the extrinsic parameters and $C_X$ is the $5 N_\Sig \times 4$ cross-correlation matrix that mixes the extrinsic and intrinsic parameters.
The Fisher matrix can similarly be decomposed:
\begin{equation}
\Gamma = \left( \begin{array}{cc}
\Gamma_\intp  & \Gamma_X  \\
\Gamma_X^T & \Gamma_\extp \end{array} \right)
\end{equation}
Now, for partitioned symmetric matrices we have (see page 46 of the Matrix Cookbook~\cite{Petersen06thematrix})
\begin{equation}
\det C = \frac{\det C_\extp}{\det\Gamma_\intp} \, ,
\end{equation}
which implies that the volume of the posterior factors into extrinsic and intrinsic pieces, where the intrinsic part has
exactly the same form as for the glitch model:
\begin{eqnarray}
V_\Sig &=& (2 \pi)^{D/2} \sqrt{\det C} \nonumber \\
&=& (2 \pi)^{5 N_\Sig/2+2} \sqrt{ \det C_\extp} \, \prod_{i=1}^{N_\Sig}  \left(\frac{\sqrt{2} Q_i}{\pi\, {\rm SNR}_i^{5}}\right)
\end{eqnarray}

Putting all the pieces together we have 
\begin{widetext}
\begin{eqnarray}
\ln \Bay_{\Sig,\Gli} &=& \left(\frac{5N_\Sig}{2} +2 - \frac{5N_\Gli}{2} \right) (1+\ln(2\pi))+\sum_{n =1}^{N_\Sig} \ln\left( \frac{\sqrt{2} Q_n}{\pi\, {\rm SNR}_n^{5}}\right)  -  \ln  p( \intp_{\rm MAP}|\Gli) \nonumber \\
&+& \ln\left( \frac{ \sqrt{ \det C_\extp}}{4\pi^2}\right) - \sum_{n =1}^{N_\Gli} \ln\left( \frac{\sqrt{2} Q_n}{\pi\, {\rm SNR}_n^{5}}\right) +  \ln  p( \intp_{\rm MAP}|\Sig)
\end{eqnarray}
\end{widetext}

From here we can insert the SNR scalings for $N_\Sig$, $N_\Gli$ and the ${\rm SNR}_n$ and include the explicit expression for the intrinsic parameter volumes in an effort to
make quantitative predictions. While the expressions are more complicated than the aligned collocated case the scalings with SNR are the same.

We are unable to derive an analytic expression for $\sqrt{ \det C_\extp}$. Additionally, there is the problem that some of the extrinsic parameters, most notably the
ellipticity and polarization angle, are poorly constrained and Fisher matrix estimates are unreliable.  As a result the posterior volume does not scale as ${\rm SNR}^{-4}$ as we naively expect
from the Fisher matrix, but as some lower power such as ${\rm SNR}^{-2}$ or ${\rm SNR}^{-3}$. One way to incorporate the restriction that the posterior not exceed the
prior is to elevate the extrinsic parameters from their fundamental domain (with periodic boundary conditions) to the universal cover, and introduce a Gaussian prior
on the parameters that restricts the posterior volume to be no larger than the prior volume. The negative Hessian of second derivatives of the log of this prior is added
to the Fisher matrix (so that the Fisher matrix describes the curvature of the posterior, not just the likelihood). Numerically computing the posterior volume as a function of
SNR using this approach shows that the posterior volume scales as  ${\rm SNR}^{-\gamma}$, where the exponent $\gamma$ is weakly dependent on the SNR, varying
between roughly 2 and 3 across the range of SNRs we expects to encounter, as shown in Fig.~\ref{fig:prediction_laplace}.

\begin{figure*}
\begin{center}
\mbox{
\includegraphics[width=\linewidth]{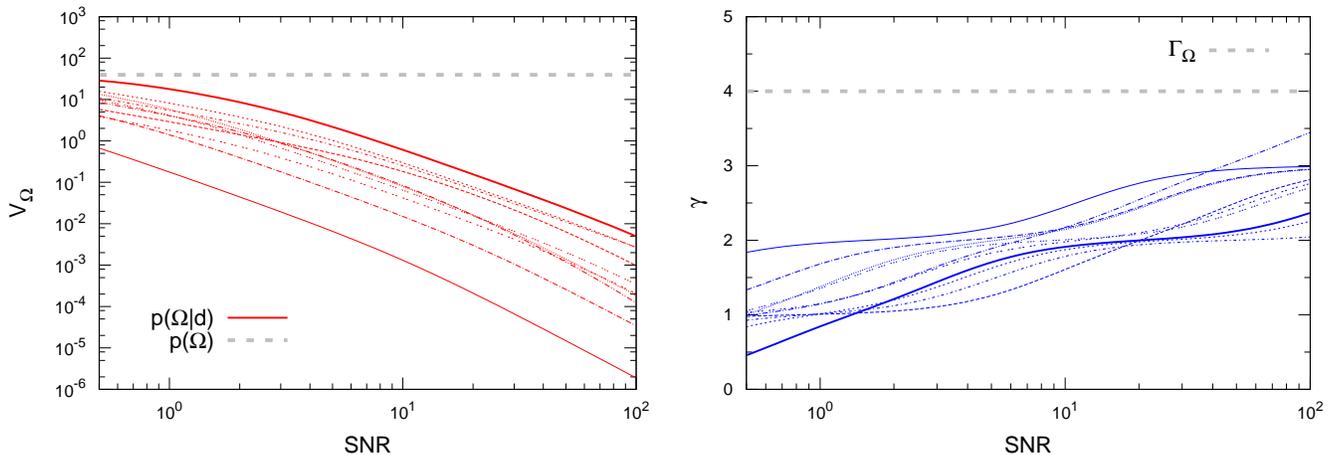} 
}
\caption{The panel on the left shows the scaling of the extrinsic posterior volume with SNR for ten randomly chosen single sine-Gaussians in a two detector network. The panel on the
right shows the SNR dependence of the slope parameter $\gamma$.}
\label{fig:extrinsic_volume}
\end{center}
\end{figure*}

The end result is that including the intrinsic parameters increases the dimension of the signal model by between 2 and 3 degrees of freedom, not 4 as we would naively expect. Thus, the overall scaling
for the single sine-Gaussian Bayes factor should scales as $\ln \Bay_{\Sig,\Gli} \sim  (5 \rightarrow 6) \ln {\rm SNR}$. The scaling for more elaborate waveforms is far more complicated.  Ultimately it is this added complexity that enables \BayesWave{} to assign high-confidence to detection candidates of non-trivial GW signals.

\end{document}